\documentclass[preprint2]{aastex}
\usepackage{natbib}

\shortauthors{Wright \etal}

\shorttitle{WISE Mission}

\newcommand{\etal}         {{\it et al.}}
\newcommand{\vs}           {{\it vs.}}
\newcommand{\asec}	{\mbox{$^{\prime\prime}$}}

\newcommand{\be}           {\begin{equation}}
\newcommand{\ee}           {\end{equation}}
\newcommand{\bea}          {\begin{eqnarray}}
\newcommand{\eea}          {\end{eqnarray}}

\begin{document}

\slugcomment{draft \today}

\title{The Wide-field Infrared Survey Explorer (WISE): Mission 
Description and Initial On-orbit Performance} 
\author{
Edward L.\ Wright\altaffilmark{1},
Peter R.\ M.\ Eisenhardt\altaffilmark{2},
Amy Mainzer\altaffilmark{2},
Michael E.\ Ressler\altaffilmark{2},
Roc M.\ Cutri\altaffilmark{3},
Thomas Jarrett\altaffilmark{3},
J.\ Davy Kirkpatrick\altaffilmark{3},
Deborah Padgett\altaffilmark{3},
Robert S.\ McMillan\altaffilmark{4},
Michael Skrutskie\altaffilmark{5},
S.\ A.\ Stanford\altaffilmark{6,7},
Martin Cohen\altaffilmark{8},
Russell G.\ Walker\altaffilmark{8},
John C.\ Mather\altaffilmark{9},
David Leisawitz\altaffilmark{9},
Thomas N.\ Gautier III\altaffilmark{2},
Ian McLean\altaffilmark{1},
Dominic Benford\altaffilmark{9},
Carol J.\ Lonsdale\altaffilmark{10},
Andrew Blain\altaffilmark{13},
Bryan Mendez\altaffilmark{16},
William R.\ Irace\altaffilmark{2},
Valerie Duval\altaffilmark{2},
Fengchuan Liu\altaffilmark{2},
Don Royer\altaffilmark{2},
Ingolf Heinrichsen\altaffilmark{2},
Joan Howard\altaffilmark{11},
Mark Shannon\altaffilmark{11},
Martha Kendall\altaffilmark{11},
Amy L.\ Walsh\altaffilmark{11},
Mark Larsen\altaffilmark{12},
Joel G.\ Cardon\altaffilmark{12},
Scott Schick\altaffilmark{15},
Mark Schwalm\altaffilmark{17},
Mohamed Abid\altaffilmark{2},
Beth Fabinsky\altaffilmark{2},
Larry Naes\altaffilmark{14},
Chao-Wei Tsai\altaffilmark{3}
}

\altaffiltext{1}{UCLA Astronomy, PO Box 951547, Los Angeles CA 90095-1547}
\altaffiltext{2}{Jet Propulsion Laboratory, 4800 Oak Grove Drive, Pasadena, CA 91109}
\altaffiltext{3}{Infrared Processing and Analysis Center,
California Institute of Technology, Pasadena CA 91125}
\altaffiltext{4}{University of Arizona, 1629 E. University Blvd., Tucson, AZ 85721, USA}
\altaffiltext{5}{Department of Astronomy, University of Virginia, Charlottesville VA 22903}
\altaffiltext{6}{Physics Department, University of California, Davis, CA 95616}
\altaffiltext{7}{Institute of Geophysics and Planetary Physics,
LLNL, Livermore, CA, 94551}
\altaffiltext{8}{Monterey Institute for Research in Astronomy, 200 8th
Street, Marina CA 93933}
\altaffiltext{9}{NASA Goddard Space Flight Center, Greenbelt MD 20771}
\altaffiltext{10}{National Radio Astronomy Observatory, Charlottesville VA 22903}
\altaffiltext{11}{Ball Aerospace \& Technologies Corporation, 1600 Commerce Street,
Boulder, CO 80301}
\altaffiltext{12}{Space Dynamics Laboratory, 1695 North Research Park Way,
North Logan, UT 84341}
\altaffiltext{13}{California Institute of Technology, Pasadena CA 91125}
\altaffiltext{14}{Lockheed Martin Advanced Technology Center (retired), Palo Alto, CA}
\altaffiltext{15}{Practical Technology Solutions, Inc., P.O Box 6336, North Logan, Utah 84341}
\altaffiltext{16}{Space Sciences Laboratory, University of California, Berkeley CA 94720}
\altaffiltext{17}{L-3 Communications  SSG-Tinsley,  Wilmington, MA 01887}

\email{wright@astro.ucla.edu}
\begin{abstract}
The all sky surveys done by the Palomar Observatory Schmidt, the European Southern 
Observatory Schmidt, and the United Kingdom Schmidt, 
the InfraRed Astronomical Satellite and the 2 Micron All Sky Survey
have proven
to be extremely useful tools for astronomy with value that lasts for
decades.
The Wide-field Infrared Survey Explorer is mapping the whole sky
following its launch on 14 December 2009.  WISE began surveying the sky on 
14 Jan 2010 and completed its first full coverage of the sky on July 17.
The survey will continue to cover the sky a second time until 
the cryogen is exhausted (anticipated in November 2010).
WISE is achieving 5 sigma point source sensitivities better than 0.08, 0.11, 1 and 6 mJy in
unconfused regions on the ecliptic in bands centered at wavelengths of 3.4, 4.6, 12 \& 22 $\mu$m.
Sensitivity improves toward the ecliptic poles due to denser coverage and lower zodiacal background.
The angular
resolution is 6.1\asec,  6.4\asec,  6.5\asec\ \& 
12.0\asec\  at 3.4, 4.6, 12 \& 22 $\mu$m, and the astrometric precision for high SNR sources 
is better than $0.15\asec$.
\end{abstract}
\maketitle
\section{Introduction}

The Wide-field Infrared Survey Explorer (WISE) will complete a
mid-infrared survey of the entire sky by mid-July 2010 with much
higher sensitivity than previous infrared survey missions.  The most 
comparable previous mission,
the InfraRed Astronomical Satellite \citep[IRAS;][]{neugebauer/etal:1984, beichman/etal:1988}
was launched in 1983 and mapped the whole sky in 4 bands
with 62 individually wired detectors.  
The AKARI mission \citep{murakami/etal:2007}
used an ingeneous technique to survey the mid-IR sky at 9 \& 18 $\mu$m
with sensitivities of 50 \& 100 mJy \citep{ishihara/etal:2010} 
with better angular resolution than IRAS.
WISE is mapping the whole sky in 4 infrared bands
W1..W4 centered at 3.4, 4.6, 12 \& 22 $\mu$m using a 40 cm telescope feeding
arrays with a total of 4 million pixels.  
The increased number of detectors leads to a 
much higher sensitivity: WISE is achieving a sensitivity more than one hundred 
times better than IRAS in the 12 $\mu$m band.
While IRAS had two far-infrared bands at 60 and 100 $\mu$m,
WISE has two mid-infrared bands at 3.4 \& 4.6 $\mu$m.  In
these bands the only all-sky survey that has been done is from the
Diffuse InfraRed Background Experiment (DIRBE) on the COBE satellite,
and DIRBE used only a single pixel per band observing a
$0.7^\circ$ beam.  A point source catalog was constructed by
\citet{smith/price/baker:2004} from the DIRBE data
with flux limits of 60 \& 50 Jy at 3.5 \& 4.9 $\mu$m, and WISE
should reach flux limits that are $5\times10^4$ times lower in these
bands.

\section{Implementation}

\begin{figure}[tbp]
\plotone{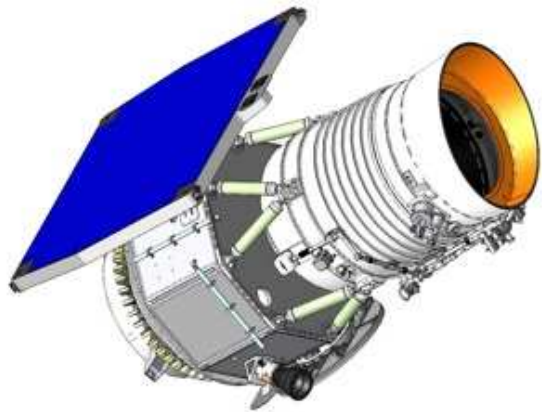}
\caption{
Diagram showing the WISE flight system in survey configuration with
cover off.  The spacecraft bus to the left of the bipod supports was 
provided by BATC, and the cryogenic instrument to the right of the bipods
was provided by SDL.
\label{fig:WISE-cover-off}}
\end{figure}

\begin{figure}[t]
\plotone{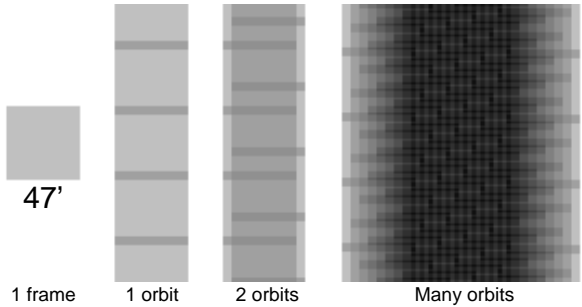}
\caption{Coverage by WISE in one frame (every 11 seconds), in one orbit,
in two orbits, and in many orbits.  The gray levels show the depth of
coverage with the darker areas having more coverage.. \label{fig:scanpat5}}
\end{figure}

\begin{figure}[t]
\plotone{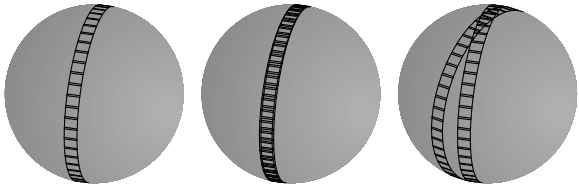}
\caption{Cartoon showing WISE coverage on the sphere for 1 orbit, for 2 consecutive orbits,
and for 2 orbits separated by 20 days, illustrating the highly redundant coverage at the
ecliptic poles.
\label{fig:scanpat1a}}
\end{figure}

WISE is a MIDEX (medium class Explorer) mission funded by NASA.
The project is managed and operated by the Jet Propulsion Laboratory (JPL) for
the PI, Edward L. Wright.  Major components of the project were
built by contractors, with the Space Dynamics Laboratory (SDL) of
Utah State University building the instrument and Ball Aerospace
\& Technologies Corporation building the spacecraft.  
See \citet{liu/etal:2008} for more details about the development of
the WISE flight system.
The data processing
and analysis is being done by the Infrared Processing and
Analysis Center (IPAC).  Education and Public Outreach activities for WISE are led by the
UC Berkeley Space Sciences Lab \citep{mendez:2008}.

The WISE flight system is 285 cm tall, 200 cm wide and 173 cm deep.
It has a mass of 661 kg.
It uses 301 W of power, while the solar panels can provide over 500 W.
The 40 cm diameter telescope sits inside a solid-hydrogen cooled
cryostat.  The cryostat plus telescope and camera have a mass of 347 kg,
and the solid hydrogen has a mass of 15.7 kg at launch.  
Figure \ref{fig:WISE-cover-off} shows the flight system in operational
configuration.

\subsection{Mission Design}

\begin{figure}[tbp]
\plotone{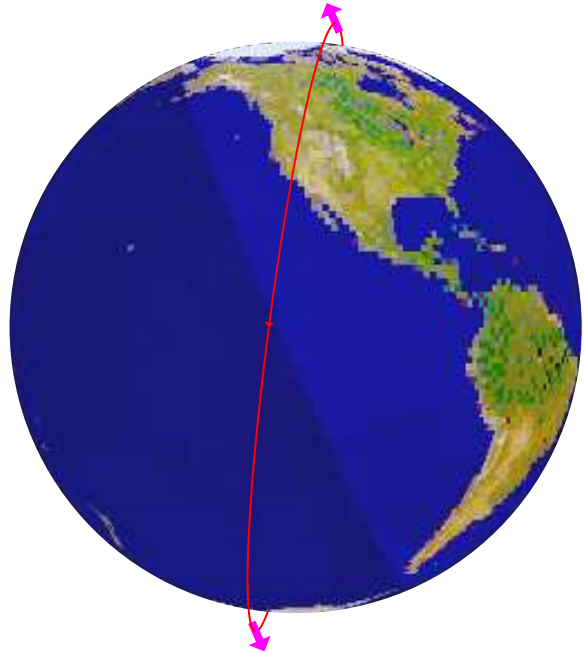}
\caption{
WISE pointing and orbit during the June solstice.  Note that WISE points
perpendicular to the Earth-Sun line and not toward the zenith.
\label{fig:Summer_solstice}}
\end{figure}

\begin{figure}[tb]
\plotone{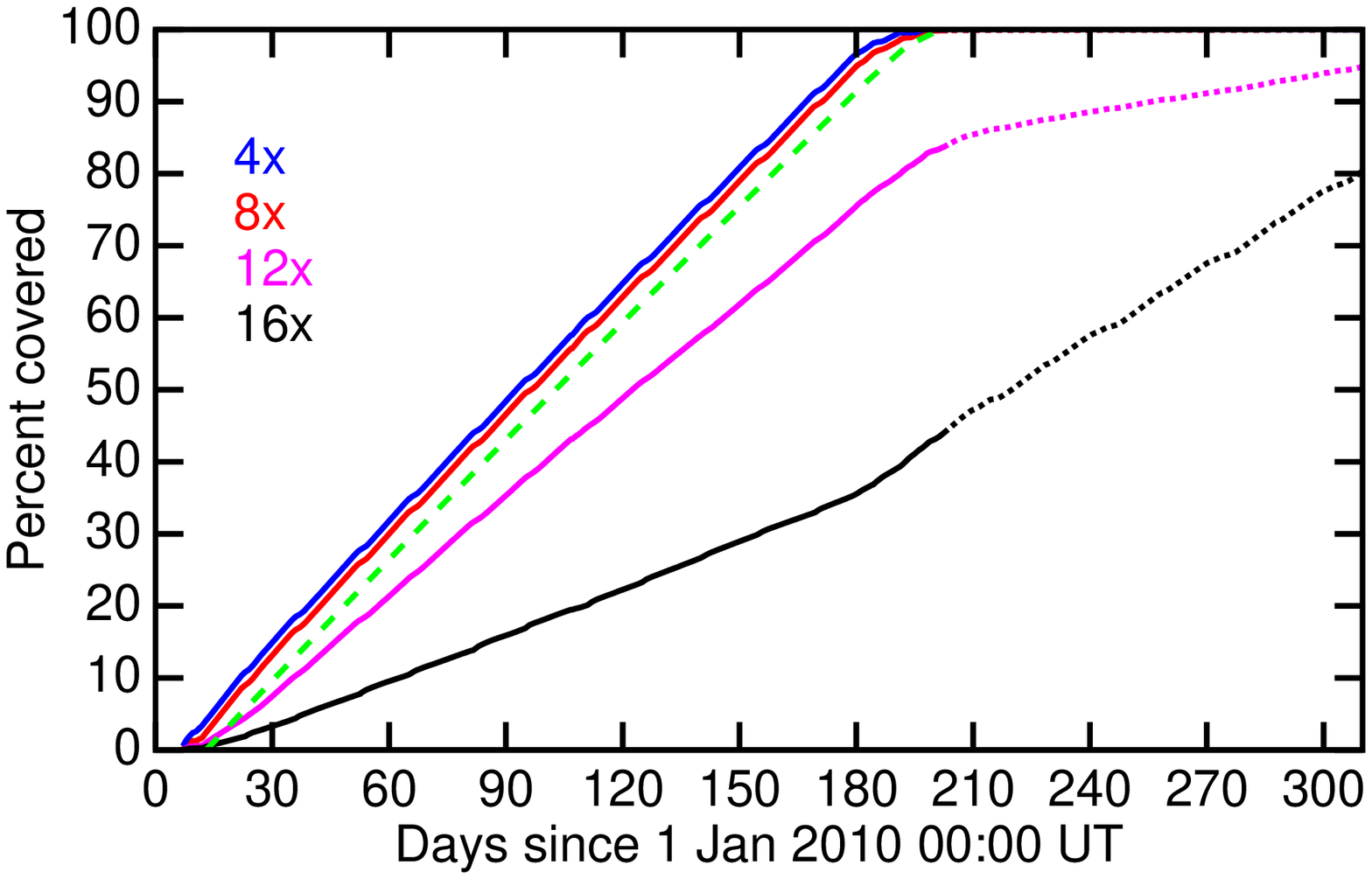}
\caption{
WISE survey coverage {\it vs.} date.  As of 15 July 2010 more than
99\% of the sky had been covered to depth of 8 frames or more.
The fraction of the sky covered to a depth of 4, 12 or 16 frames is also
shown.  
The dotted portions of the curves show the anticipated coverage
until the expected exhaustion of the cryogen.
The vertical dashed line shows when the 22 $\mu$m channel
saturated after the actual exhaustion of the secondary tank.
The small slowdowns in the survey progress occur when
the Moon crosses the scan path.
The dashed line shows a simple prediction of the sky
coverage based only on the longitude of the Sun, starting on
14 Jan 2010.
\label{fig:progress}}
\end{figure}

It is easiest to observe sources when the Sun is not in the same part of the sky,
since maximizing the elongation
from the Sun makes it much easier to keep optics cool, and to avoid
scattered light from the Sun.  
But to scan the entire sky one must observe the ecliptic poles,
which are always 90$^\circ$ from the Sun.  Thus WISE adopted a basic
strategy of scanning great circles with a center located at the Sun,
which keeps the solar elongation at 90$^\circ$.  These circles scan from
the North Ecliptic Pole (NEP) to the South Ecliptic Pole (SEP),
crossing the ecliptic at longitudes $\pm90^\circ$ from the Sun.  The entire
sky is covered by this scan pattern  in half a year as the Earth-Sun line turns by
180$^\circ$.

Confusion noise would prevent detection of key WISE science targets unless 
the beam size were less than 50 square arc-seconds,  and since  there are $5 \times 
10^{11}$ square arc-seconds in the sky, WISE clearly needs to
transmit large quantities of data to the ground.  
This led to choosing a low Earth orbit to minimize the transmission loss.  The
actual WISE orbit started with a mean radius of 6909 km in mid-December
2009, and the radius had decreased by 0.8 km due to atmospheric drag by mid-June 2010.  
In a polar orbit the average geoid height is $(R_{eq}+R_p)/2 = 6368$~km,
so the average altitude above the geoid is 540
km.  This altitude was chosen to reduce the exposure to trapped radiation belts in the 
South Atlantic Anomaly (SAA).  
Since looking at the Earth
would both blind the detector arrays and overwhelm the cryogenic system, 
a Sun synchronous orbit was chosen, with an inclination
(97.5$^\circ$) chosen to give a precession rate of 360$^\circ$ in
one year.  The right ascension of the ascending node was chosen so
that the local time at the equator crossings is either 6 AM or 6
PM.  For the actual WISE launch at 14:09:33 UT on 14 Dec 2009 the ascending node
is at 6 PM local time.  Since the end of the orbit normal close to the Sun is 
at $-7.5^\circ$ declination for dawn launch from the Western Test Range (at Vandenberg AFB, CA), during
June the angle between Sun and the orbit normal can be as
large as $31^\circ$, and eclipses occur over the South pole.
COBE was in a 99$^\circ$ inclination
orbit at 900 km altitude with a June-centered eclipse season.
The alternative of a 6 PM launch from the WTR gives an
eclipse season centered on the December solstice, as was the case for IRAS.

By synchronizing the scans and the orbit so the telescope is pointed at the NEP
when the satellite is closest to the North pole of the Earth, one achieves a scan
pattern that covers the entire sky, with the line of sight (LOS) never closer to the
Sun than 90$^\circ$ and never further than 31$^\circ$ from the zenith.

WISE has adopted the
2MASS strategy of moving the telescope at a constant inertial rate,
and using a scan mirror to freeze the sky on the arrays for the 
integration time.  The field of view (FOV) of WISE is 47$^\prime$, and by choosing to have
a small (10\%) overlap between frames, the cadence between frames is
set at 11.002 seconds per frame.  For WISE 8.8 seconds of this is spent
integrating and it takes about 1.06 seconds to read out the arrays.  
The LOS then jumps forward 42$^\prime$ in 1.1 seconds.  
There are about 15 orbits per day, so the scan circle
advances by about 4$^\prime$ per orbit, giving about 12 orbits where
the FOV hits a given source.  Figure \ref{fig:scanpat5} shows how the
redundancy builds up, while Figure \ref{fig:scanpat1a} shows scans on the
celestial sphere.
Figure \ref{fig:Summer_solstice} shows the the relationship between the
WISE orbit, the Earth-Sun line, and WISE line-of-sight during the eclipse
season.

The Moon crosses the scan circle twice a month, and this would leave gaps in
the sky coverage because the Moon is bright enough to cause stray light
problems when it is less than 15$^\circ$ from the line of sight.  Thus WISE
uses a modified scan pattern where the scan circle gets slightly ahead before
the Moon interferes, and then drops slightly behind to recover the region
the Moon blotted out.  The Moon moves 13$^\circ$ per day in ecliptic longitude,
so with a 15$^\circ$ exclusion zone WISE needs to be 1.2$^\circ$ ahead just
before the Moon crosses the scan circle, and then drop back to 
1.2$^\circ$ behind.  In addition, the SAA affects part of 4 orbits per day,
and to mitigate reductions in the sky coverage due to the SAA, a small toggle
of $\pm0.2\deg$ is added to the longitude of the scan circle center,
with the sign changing with a two orbit period.  Finally, the scan circle
central longitude is biased by $+5^\circ$ on the side of the scan where
the Sun is approaching the scan, which is the 6 PM side of the orbit.
This means that the scans are actually 90$^\circ$ and 95$^\circ$ from the
Sun when crossing the ecliptic, allowing for a 5 day recovery period
after a satellite anomaly without causing a gap in sky coverage.

The WISE mission design provides at least 8 frames on over 99\% of the
sky in a 6 month survey interval
after allowing for data lost to the Moon and the SAA.
\citet{heinrichsen/wright:2006} describe the mission
operations system in more detail.
Figure \ref{fig:progress} shows the survey progress {\it vs.}  date.
In mid-July 2010 WISE
completed its first pass over the sky, but will continue to survey until its
cryogen runs out.

\citet{duval/etal:2004} and \citet{mainzer/etal:2005} describe the WISE hardware design.

\subsection{Instrument}

\begin{figure}[tb]
\plotone{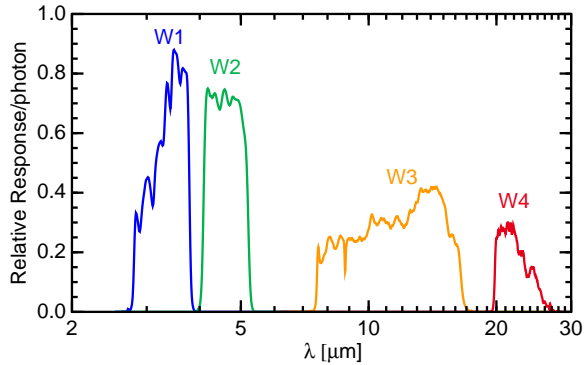}
\caption{The weighted mean WISE relative spectral response functions
in electrons per photon.
\label{fig:RSR-best-lin}}
\end{figure}

\begin{figure}[tb]
\plotone{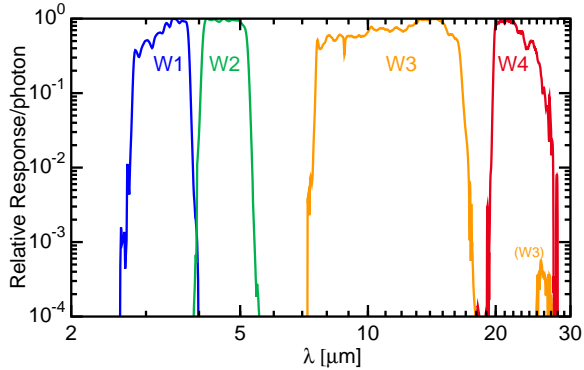}
\caption{The weighted mean WISE relative spectral response functions
after normalizing to a peak value of unity, on a logarithmic
scale.
\label{fig:RSR-best-log}}
\end{figure}

The WISE telescope, imager, dichroic beamsplitter, detectors, and cryostat
were built by the Space Dynamics Laboratory of Utah State
University with major subcontracts from Lockheed Martin Advanced
Technology Center, L3 Communications SSG-Tinsley, and DRS Technologies.  
\citet{larsen/schick:2005} describe
the WISE science payload in more detail.  The instrument was tested,
then passed a vibration test, and was tested again, before being delivered
to Ball Aerospace in May 2009 for final integration with the spacecraft.
The testing included a determination of the optimal focus, a measurement
of the point spread function, and a determination of both the relative
spectral response and the absolute sensitivity of the system.

The WISE short-wavelength channels employ 4.2 and 5.4 $\mu$m cutoff
HgCdTe arrays fabricated by 
Teledyne Imaging Sensors with $1024\times1024$ pixels each 18 $\mu$m square. 
For the long-wavelength channels, the detectors are Si:As BIB arrays from DRS Sensors \& Targeting Systems
with the same $1024\times1024$ pixel format and pitch.
The first four and last four pixels in each row and column are used as
non-illuminated reference pixels, so the effective size of the arrays is
$1016 \times 1016$ pixels.  The median pixel scale is 2.757\asec/pixel
with a range of $\pm 0.6\%$ among the different axes and arrays.  
The band 4 data is binned $2\times2$ on board, giving 5.5\asec/pixel.
All four arrays image the same field of view simultaneously using 3 dichroic beam
splitters.

All of the arrays use sampling up the ramp.  The 11 sec cadence between
frames is divided into 10 parts.  A reset-read of the arrays occurs during the
first 1.1 sec, followed by 8 read cycles each 1.1 sec apart and a final
reset cycle while the scan mirror flies back.  For the W3 and W4 arrays the 
9 reads are multiplied by weights of -4, -3, -2, -1, 0, 1, 2, 3 \& 4 and summed
to give the slope of the ramp.  The W1 and W2 arrays showed excess noise
in the first sample so the weights used are 0, -7, -5, -3, -1, 1, 3, 5 \& 7.

The long-wavelength arrays show long lasting latent images in the
flat-field which are removed by annealing these arrays twice per
day.  The annealing heaters are on for 90 seconds and heat the
arrays from their 7 K normal operating temperature to 15 K, which
removes the latent images.

The spectral response of the system has been determined in three
different ways:  the whole system response was measured using a
Fourier transform spectrometer (FTS), the system response was
estimated using the product of component data (measured with an
FTS), and the response was computed from component design calculations.
Since
the sharp edges in the response function cause ringing and negative
values in the FTS spectrum, a ``best'' combined estimate of the relative spectral
response was computed as follows:  first a power law was fit to the
ratio of the total system measured response to either the component
prediction or the design prediction.  The predictions were then
adjusted to better match the measured transmission using these fits.
The fits were weighted by the square of the response function times
a $1/\sigma^2$ statistical weight, so the correction is best in the
filter passbands.  After the adjustment the three methods gave
consistent responses.  
Finally the best estimate for the spectral response was made using
a weighted mean of the measured, predicted from components, and
design values.  In the weighted mean, zero or negative responses
were given zero weight, the design and component predictions were
given unit weight, and the measurements were given weights of
SNR$^2$.  This weighted mean relative spectral response is shown
in Figures \ref{fig:RSR-best-lin} and \ref{fig:RSR-best-log}.  The
curves in the logarithmic plot
have been normalized to a peak value of 1.  Note that the
small red leak in the W3 transmission is based on data with a low
SNR and may not actually exist.

WISE measures the signal $S$ given by
\be
S \propto \int R(\lambda) F_\nu d\ln\nu 
\propto \int R(\lambda) \lambda F_\lambda d\lambda
\ee
since the relative spectral response $R$ is electrons per photon and $d n_\gamma \propto
F_\nu d\nu/(h\nu) \propto F_\nu d\ln\nu$. 

\begin{deluxetable}{lrrrrrrr}
\tablecolumns{5}
\tablecaption{\label{tab:colorcorr} Flux corrections and colors for powerlaws
and blackbodies}
\tablehead{
\colhead{$F_\nu$} & \colhead{$f_c(W1)$}  & \colhead{$f_c(W2)$} & \colhead{$f_c(W3)$} & 
\colhead{$f_c(W4)$} & \colhead{[W1-W2]} & \colhead{[W2-W3]} & \colhead{[W3-W4]} }

\startdata
\hline
$\nu^3$ &  1.0283 &  1.0206 &  1.1344 &  1.0142 & -0.4040 & -0.9624 & -0.8684 \\
$\nu^2$ &  1.0084 &  1.0066 &  1.0088 &  1.0013 & -0.0538 & -0.0748 & -0.0519 \\
$\nu^1$ &  0.9961 &  0.9976 &  0.9393 &  0.9934 &  0.2939 &  0.8575 &  0.7200 \\
$\nu^0$ &  0.9907 &  0.9935 &  0.9169 &  0.9905 &  0.6393 &  1.8357 &  1.4458 \\
$\nu^{-1}$ &  0.9921 &  0.9943 &  0.9373 &  0.9926 &  0.9828 &  2.8586 &  2.1272 \\
$\nu^{-2}$ &  1.0000 &  1.0000 &  1.0000 &  1.0000 &  1.3246 &  3.9225 &  2.7680 \\
$\nu^{-3}$ &  1.0142 &  1.0107 &  1.1081 &  1.0130 &  1.6649 &  5.0223 &  3.3734 \\
$\nu^{-4}$ &  1.0347 &  1.0265 &  1.2687 &  1.0319 &  2.0041 &  6.1524 &  3.9495 \\
$B_\nu( 100)$ & 17.2062 &  3.9096 &  2.6588 &  1.0032 & 10.6511 & 18.9307 &  4.6367 \\
$B_\nu( 141)$ &  4.0882 &  1.9739 &  1.4002 &  0.9852 &  7.7894 & 13.0371 &  3.4496 \\
$B_\nu( 200)$ &  2.0577 &  1.3448 &  1.0006 &  0.9833 &  5.4702 &  8.8172 &  2.4949 \\
$B_\nu( 283)$ &  1.3917 &  1.1124 &  0.8791 &  0.9865 &  3.8329 &  5.8986 &  1.7552 \\
$B_\nu( 400)$ &  1.1316 &  1.0229 &  0.8622 &  0.9903 &  2.6588 &  3.8930 &  1.2014 \\
$B_\nu( 566)$ &  1.0263 &  0.9919 &  0.8833 &  0.9935 &  1.8069 &  2.5293 &  0.8041 \\
$B_\nu( 800)$ &  0.9884 &  0.9853 &  0.9125 &  0.9958 &  1.1996 &  1.6282 &  0.5311 \\
$B_\nu(1131)$ &  0.9801 &  0.9877 &  0.9386 &  0.9975 &  0.7774 &  1.0421 &  0.3463 \\
K2V &  1.0038 &  1.0512 &  1.0030 &  1.0013 & -0.0963 &  0.1225 & -0.0201 \\
G2V &  1.0049 &  1.0193 &  1.0024 &  1.0012 & -0.0268 &  0.0397 & -0.0217 \\
\hline
\enddata
\end{deluxetable}

Specifying a central wavelength for broad bands like the WISE filters,
especially the very wide W3 band, is always ambiguous.  We have chosen to
specify the isophotal wavelengths for our filters. 
The isophotal wavelength and magnitude zeropoint are defined
by requiring that a constant $F_\lambda^\circ$ give the same
signal as Vega in a WISE band, and that
$F_\lambda^\circ = F_\lambda^{Vega}(\lambda_{iso})$.
\citet{tokunaga/vacca:2005} give a clear definition of the
isophotal wavelength, and values for Mauna Kea Observatory
filter set.  
\citet{cohen/etal:1992} give the IR spectrum of Vega, based on a
model from Kurucz (1991, private communication to Martin Cohen).
It is important to remember that WISE saturates on Vega, so
the magnitude zeropoint is actually based on fainter stars calibrated to
the Vega system.  Therefore the debris disk around Vega \citep{aumann/etal:1984} is relevant
only if it has affected the calibration used to determine the
magnitudes of the fainter standards.
A convenient and accurate fit to the continuum of our Vega spectrum in the
$2.5 < \lambda < 29\;\mu$m range we use is given by
\bea
F_\lambda & =  & 1.0158 \times 10^{-16}
(1-0.0083\ln(\lambda/8.891\,\mu\mbox{m})^2)
\nonumber\\
& & \times B_\lambda(14454\;\mbox{K}),
\eea
where $B_\lambda$ is the Planck function,
which matches the continuum spectrum to an average absolute error of 0.045\%.
The temperature in this ad hoc fit is much higher than the effective temperature of Vega
and the solid angle is much lower than the real solid angle due to the increase with
wavelength of the free-free and bound-free opacites.
Absorption lines reduce the in-band fluxes by only 0.73, 0.57, 0.28 \& 0.14\% in bands 1 through 4.
To avoid possible multiple solution we use the continuum spectrum when solving for $\lambda_{iso}$.
The isophotal wavelengths so defined are
3.3526,  4.6028, 11.5608 \& 22.0883 $\mu$m for W1..4. 
The measured system transmissions were smaller than the expected values at
long wavelengths, leading to an effective wavelength for WISE band 4 of
22 $\mu$m instead of the expected 23 $\mu$m.
Absolute measurements of Vega by MSX  \citep{price/etal:2004} 
show that a 2.7\% upward offset from the model spectrum is needed 
at 21.3 $\mu$m \citep{cohen:2009}.  We have applied this
correction to the WISE 22 $\mu$m band, giving
magnitude zeropoints on the Vega system in the WISE passbands of
$F_\lambda^\circ = 8.180\times10^{-15},\; 2.415\times10^{-15},\;  
6.515\times10^{-17}\;\&\;5.090\times10^{-18}$
W/cm$^2$/$\mu$m, which convert to 
$F_\nu^\circ = 306.681,\;170.663,\;29.0448\;\&\;8.2839$ Jy in
W1..W4 using 
$F_\nu^\circ = (\lambda_{iso}^2/c) F_\lambda^\circ$.
There is an overall systematic uncertainty of $\pm 1.5\%$ from the 
Vega spectrum in these flux zeropoints.

We have also observed in-flight a discrepancy between red
(typically sources with $F_\nu \propto \nu^{-2}$) and blue
(stars with $F_\nu \propto \nu^2$)
calibrators in W3 and W4, the 12 and 22 $\mu$m bands.  This amounts to
about -17\% and 9\% in the fluxes for W3 and W4, with the red sources
appearing too bright in W4 and too faint in W3.  The flux differences
could be resolved by adjusting the effective wavelength of W3 and W4
3-5\% blueward and 2-3\% redward, respectively.  This would change the
zero magnitude $F_\nu^\circ$ by about -8\% in W3 and +4\% in W4.
But the zeropoints and isophotal wavelengths reported here are based on
the relative spectral responses derived from ground calibration without any such
adjustment.
The instrumental zeropoints that define the conversion from counts to
magnitudes have been based on standard stars, which are the blue
calibrators.
Given the discrepancy between red and blue calibrators we estimate that
the conversion from magnitudes to Janskys are currently uncertain
by $\pm10\%$ in W3 and W4.  Updated values will be provided in
the Explanatory Supplement accompanying the preliminary data
release.

This definition of the isophotal wavelength and flux zeropoint
means that the color correction term for a source with a different
spectrum than Vega vanishes by construction when
$F_\lambda$ is a constant ($F_\nu \propto \nu^{-2}$)
and very nearly vanishes for Rayleigh-Jeans sources with 
$F_\lambda \propto \lambda^{-4}$ or $F_\nu \propto \nu^2$.
These spectral energy distributions bracket the vast majority
of WISE sources, so the the color corrections are generally small.
But the extremely wide W3 filter does lead to color corrections
as large as 0.1 magnitude for a constant $F_\nu$.
Table \ref{tab:colorcorr} gives the flux correction factors in the
WISE bands for several input spectra, and the WISE colors
for these spectra.   These factors multiply the signal $S$ given
by a spectrum $F_\nu = F_\nu^\circ (\lambda_{iso}/\lambda)^\beta$.
Thus a spectrum with $F_\nu = \mbox{const} = 29.0$ Jy gives a
signal that is $f_c(W3) = 0.9169$ times the signal from Vega, so one
would need a constant $F_\nu$  of $29.0/0.9169 = 31.7$ Jy  to
give zero magnitude.

Outer Solar system objects like Centaurs will have color temperatures
close to 100~K and for a 100~K blackbody the flux correction factor
$f_c(W3) = 2.6588$ is quite large.  Thus a 100~K blackbody would need
$29.0/2.66 = 10.9$~Jy at 11.56 $\mu$m to give zero magnitude in W3.

The most common stars in the WISE catalog at high galactic latitude should
be G-K dwarfs, so the flux correction factors and colors for a K2V and a G2V
have been included in Table \ref{tab:colorcorr} using \citet{kurucz:1993}
spectra.  Since the W2 band includes 
the fundamental CO bandhead, the flux correction factors give the ratio between the signal from
the star to the signal from a constant $F_\lambda$ equal to the average over
a log normal passband with a 9.1\% FWHM centered on $\lambda_{iso}$.

\subsubsection{Optics}

The optical design and assembly of the WISE telescope and camera
was done by L-3 Communications SSG-Tinsley.  The optics consist of an afocal 40 cm diameter
telescope that produces a parallel beam of light that is fed into the
scan mirror, which works in the parallel beam, and then into the
all-reflecting camera.   There are six mirrors including folding flats in the
afocal telescope before the scan mirror, and six mirrors in the camera after the scan
mirror.  All of the mirrors are gold-coated giving a high infrared transmission.
The design is described in more detail by
\citet{schwalm/etal:2005}.

\subsubsection{Cryostat}

The cryostat \citep{naes/lloyd/schick:2008}
was built by the Lockheed Martin Advanced Technology Center.
It uses solid hydrogen to cool the telescope to less than 12 K, and the Si:As arrays to
less than 7.5 K.  There are two tanks of solid hydrogen: a larger secondary tank that
cools the telescope and optics, the short wavelength arrays, and shields around
the smaller primary tank that just cools the Si:As long wavelength arrays.
It has a predicted on-orbit lifetime of $10.75^{+1}_{-0.5}$ months.
\citet{schick/lloyd:2009} describe the cryostat support system and test results.

However, the in-flight performance deviated from the model predictions with
the secondary tank running out of hydrogen on 5 August after 7.7 months in
orbit.  The primary tank continued to cool the long wavelength detectors,
but the telescope warmed up to 46 K producing large
backgrounds at 12 \& 22 $\mu$m.  The 22 $\mu$m channel stopped
producing useful data after 8 August 2010.  The integration time on
the 12 $\mu$m channel was cut in half on 14 August, cut in half again on
20 August, and cut in half again on 23 August.  Observations continued
in the 3.4, 4.6 \& 12 $\mu$m bands until the primary tank ran out of
hydrogen on 29 Sep 2010.

\subsection{Spacecraft}

The WISE spacecraft was built by Ball Aerospace, with a design
based on the RS300 series of single string spacecraft.  The NextSat component
of the Orbital Express mission was the first RS300 spacecraft to be
launched, and it worked flawlessly.  SWRI provided the spacecraft avionics.

WISE is a three-axis controlled spacecraft that is commanded to
follow long scans at a constant inertial rate.  The spacecraft provides attitude 
control of better than 75\asec, jitter of less than 1.3\asec, 
and drift rate variation of less than 0.2\asec/sec over 9 seconds. Angular momentum
is stored on board in 4 reaction wheels, and any buildup of excess
angular momentum is dumped using magnetic torquer rods.
Primary attitude information is provided by two Ball CT-633 star trackers.
A fiber optic inertial measurement unit, Sun sensors and magnetometers are used 
during safe and emergency modes of the spacecraft.

A fixed solar panel provides over 500 watts of power.  WISE is oriented so the solar
panel is always pointing very nearly at the Sun.  During eclipse, a 20 amp-hour 
lithium-ion battery provides flight system power.  
Science data and flight system telemetry are stored on a 96GB flash 
memory card for later transmission to earth.
The science data volume is about 50 Gbytes per day of uncompressed
data, but lossless compression reduces this by roughly a factor of 2.
A fixed high
gain antenna is used to transmit data to the Tracking and Data
Relay Satellite System (TDRSS).  
The spacecraft body steers the antenna to TDRSS for downlink.
Science operations have to be stopped
during data transmission, so the data downlinks are scheduled
while the satellite passes over the poles of the Earth, where the
sky coverage is highly redundant. 

\subsection{Mission Operations}

Except for looking up instead of down, WISE is very similar to an
Earth observing satellite.  WISE is in a Sun-synchronous low Earth
orbit like the Landsats and many other Earth observers, and it 
uses the TDRSS to download a large volume of data.  Thus WISE
is being operated by the Earth Sciences Operations Center at JPL
which also operates the JASON ocean topography mapping mission.

The preparation of a command load for WISE starts with a set  of survey
planning parameters, which then are used along with a predicted orbit
to generate the survey plan for a week in the future.  Times when a
turn to TDRSS would be compatible with the WISE Sun and Earth pointing
constraints are chosen, and a complete sequence of pointing commands
covering half a week that
combines observing with TDRSS downloads is created.  Additional commands
for momentum unloading and annealing the W3 and W4 detector arrays
are inserted into the command sequence.  This process is repeated twice
a week to allow for the uncertainties in the predicted orbit caused by the
fluctuating atmospheric drag due to solar activity.  In the first half of 2010
the solar activity has been anomalously low, minimizing the errors in the
predicted orbits.

\subsection{Data Processing \& Archiving}

WISE science data processing, archiving and distribution is performed by the 
Infrared Processing and Analysis Center (IPAC), California Institute of Technology.  
The processing software and operations system is based on algorithms, 
pipelines and architecture used for the 2MASS, Spitzer Space Telescope and GALEX projects.

Following each TDRSS downlink contact, 
science image data packets are sent to 
IPAC where they are uncompressed and assembled 
into raw FITS format images in the four WISE bands. 
The raw FITS image headers are populated with flight 
system engineering telemetry, navigation 
data and survey events information that is sent 
asynchronously from Mission Operations at JPL.  
The images from each downlink transfer are then 
passed to the WISE Scan/Frame pipeline that 
performs instrumental calibration (droop correction, 
linearization, bias subtraction, flat-fielding), 
source extraction, photometric and astrometric calibration 
and artifact identification on the individual exposure frames. 
The WISE Multiframe pipeline operates on collections of 
calibrated single-frame images that cover a region on the sky.  
The single-frame images are coadded by registering them onto 
a common pixel grid, matching local background levels, 
and optimally combining resampled pixel values using outlier 
rejection to suppress transient pixel events 
such as cosmic rays, noise excursions and fast moving 
objects \citep{masci/fowler:2009}.  
Sources are detected and characterized on both the resulting 
combined images and individual frames, 
and spurious detections of image artifacts are identified and flagged.  
Both Scan/Frame and Multiframe processing are accompanied 
by detailed quality assessment to 
monitor data accuracy relative to WISEÕs science data requirements.

Source detection and characterization in the Scan/Frame and Multiframe 
pipelines exploit WISEÕs simultaneous four band measurements.  
Sources are detected by thresholding on a combined four-band, 
matched-filter image, similar to the ``chi-square'' detection method 
described by \citet{szalay/connelly/szokoly:1999}. 
Profile-fit photometry is performed on all bands simultaneously, 
using a maximum-likelihood fitting procedure that also operates on 
multiple, independent image frames in the Multiframe pipeline.  

Astrometric calibration of single-exposure WISE images is 
performed by solving for the frame center positions and rotations using the in-situ 
measured position of 2MASS PSC astrometric reference stars.  
Band-to-band rotations and offsets, plate scales, 
and field distortion are determined by off-line analysis using the 
apparent positions of reference stars in thousands of image frames.  
Full WCS information derived from the astrometric solutions is included in the single-frame image headers.  
The single-frame astrometric solutions are propagated to the 
coadded images and extracted source lists in the Multiframe pipeline.

Processed WISE images and extracted source data and metadata will be archived 
and served to the community via the web-based and machine-friendly interfaces 
of the NASA/IPAC Infrared Science Archive (IRSA).  
WISE data products will be interoperable with other data centers and 
services through the IRSA infrastructure.

\section{On-orbit Performance}

\begin{figure}[tb]
\plotone{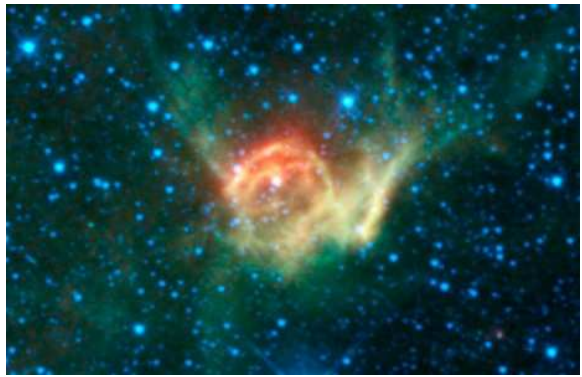}
\caption{
The Wolf-Rayet wind bubble NGC 2359 in the infrared using
W1 as blue, W2 as cyan, W3 as green, and W4 as red.
Image width is $0.4^\circ$, with North to the left and West up.
\label{fig:NGC2359}}
\end{figure}

\begin{figure}[tb]
\plotone{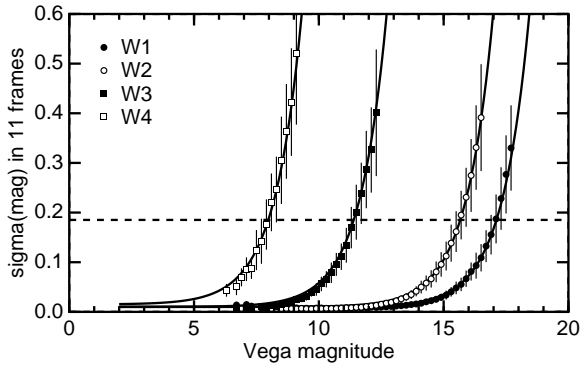}
\caption{Standard deviation of WISE magnitudes derived from the repeatability
on sources seen in 11 frames.  The standard deviations have been binned
into bins with width 0.2 mag, and the resulting histograms have been fit
with a gaussian using outlier rejection.  The points and errorbars show
the derived means and standard deviations.  The curves show a noise model
fitted to the data.  The dashed line shows an SNR of 5.86:1 which corresponds
to 5:1 in 8 frames.  From left to right,
the open squares show W4, the filled squares show W3, the open circles
show W2, and the filled circles show W1.
\label{fig:trimmed-mean-m11}}
\end{figure}

\begin{figure}[tbp]
\plotone{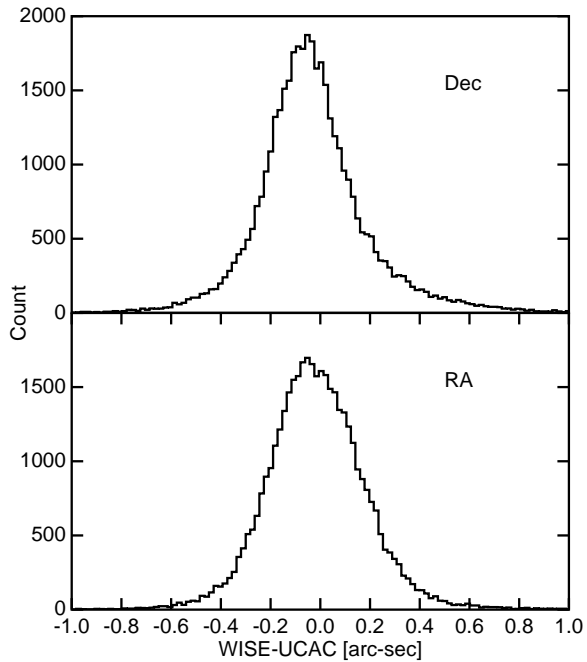}
\caption{Histograms of the difference in right ascension and declination between WISE and
UCAC positions for objects with SNR in W1 $> 20$. 
\label{fig:astrometric-scatter}}
\end{figure}

\begin{figure}[tbp]
\plotone{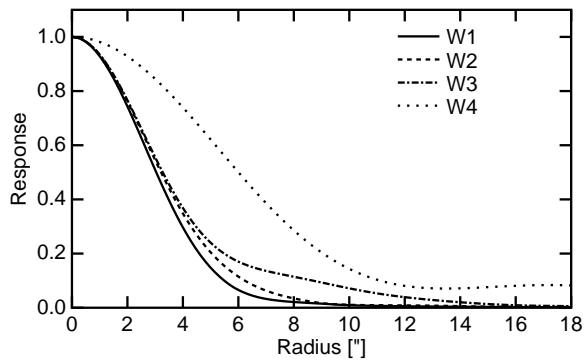}
\caption{Azimuthally averaged point spread functions for WISE.
\label{fig:psf}}
\end{figure}

WISE observes each patch of sky many times over a period of a day or more
during its normal survey operation.   These multiple frames can be combined
into co-added multicolor mosaics such as Figure \ref{fig:NGC2359}, which shows
the the Wolf-Rayet wind bubble NGC 2359 \citep{schneps/wright:1980}.
The WISE project is posting an image of the week on its multimedia gallery
which can be reached via http://wise.astro.ucla.edu.

While the sensitivity of WISE is
specified for 8 coverages, this is a worst case.  Even on the
ecliptic, where the coverage is the smallest, the most likely number of frames
covering a given source is 11. 
The individual frames are analyzed to provide astrometric and photometric
information.   Analyzing the scatter among the values from individual frames
gives the noises shown in  Figure \ref{fig:trimmed-mean-m11}.  The dashed
line shows a 5:1 SNR in 8 frames scaled to the 11 frame case plotted,
giving $\sigma(m) = 0.185$ mag
where $n$ is a noise term fitted to the data points.
The curves show
$\sigma(m) = 0.01+[2.5/\ln(10)] n/10^{-0.4m}$.
The magnitudes where the curves cross the dashed line are
17.11, 15.66, 11.40 \& 7.97 mag.
These give raw sensitivities of 44, 93, 800 \& 5500 $\mu$Jy
for $5\sigma$ in 8 frames.  There will be additional uncertainty
due to source confusion which is not included in these noises.
Allowing for confusion gives sensitivites of 0.08, 0.11, 1 \& 6 mJy.

The astrometric accuracy of WISE can be estimated by looking at the
differences between WISE and UCAC3 \citep{zacharias/etal:2010}
positions, shown in
Figure \ref{fig:astrometric-scatter} for sources with SNR $> 20$.  
At lower SNRs, an additional error
of about FWHM/(2\,SNR) must be added in quadrature.
Since this additional error is $0.15\asec$ for a PSF FWHM of $6\sec$
and SNR$=20$, and the
observed width of the distribution implies $\sigma = 0.17\asec$ for
SNR $> 20$, WISE positions for high SNR sources
should be better than $0.15\asec$ for $1\sigma$ and 1 axis.

The image quality for WISE is good, as shown in Figure \ref{fig:psf}.
These curves show the azimuthally averaged point spread function
for WISE with FWHMs of 6.1\asec,  6.4\asec,  6.5\asec\ \& 
12.0\asec\ in WISE bands 1 to 4.  In the geometric limit without
diffraction the estimated FWHM is roughly 5.5\asec.

The sensitivity of WISE to diffuse emission can be estimated
from the point source sensitivity and the beam size.   The point
source flux sensitivity scales like the square root of the noise
effective solid angle $\Omega_e$ \citep{wright:1985}, also
known as noise pixels.
Thus the diffuse sensitivity for a source solid angle
$\Omega$ is given by
$\sigma(I) = \sigma(F)/\sqrt{\Omega_e\Omega}$.
The noise pixels for W1..W4 are 13.4, 16.7,
34.3 (2.75\asec), and 25.9 (5.5\asec).  These give
$\Omega_e = 2.4, 3.0, 6.1\; \& \; 18.4 \;\mbox{nsr}$ in W1..W4.
For a source size of $5^\prime \times 5^\prime$ or
$\Omega = 2.1 \; \mbox{$\mu$sr}$, the $5\sigma$ diffuse
sensitivities from statistical noise alone are $0.6, 1.2, 7 \; \& \; 28 \; \mbox{kJy/sr}$
in W1..W4.
The sensitivity to large scale structures will also be limited by errors
in the flat field corrections and inaccuracies in matching the backgrounds
from frame to frame.

The observed saturation levels for WISE are 0.3, 0.5, 0.7 \& 10 Jy
for W1..W4.  Saturated pixel values are replaced with a
flag indicating which of the 9 samples up the ramp was the first to
saturate, so there is still a coarse indication of the brightness in saturated
regions.

\section{Science Goals}

\begin{figure}[tbp]
\plotone{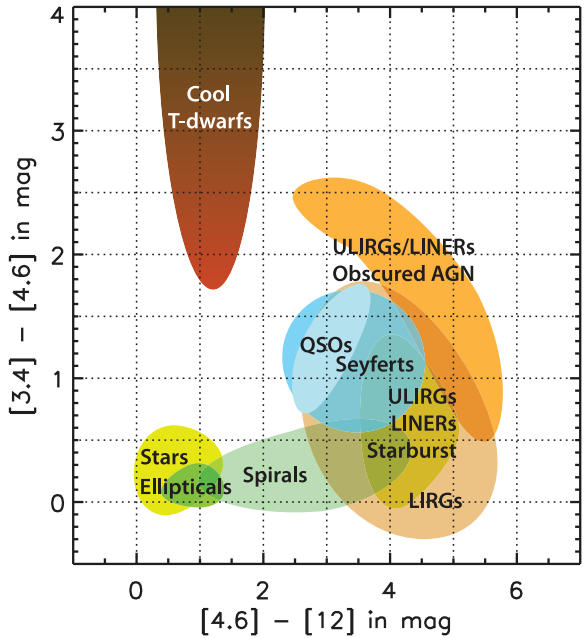}
\caption{Color-color diagram showing the locations
of interesting classes of objects.  Stars and early type galaxies
have colors near zero, while brown dwarfs are very red in W1-W2,
spiral galaxies are red in W2-W3, and ULIRGS tend to be red in both
colors.
\label{fig:bubble}}
\end{figure}

WISE  has achieved 5$\sigma$ point source sensitivities 
better than
0.08, 0.11, 1 \& 6 mJy at 3.4, 4.6, 12 \& 22 $\mu$m
in regions observed in 8 or more frames, which is expected to be more than
99\% of the sky.
These sensitivities correspond to Vega magnitudes of
16.5,  15.5, 11.2 \& 7.9.  Thus WISE will go a magnitude deeper than the 2MASS
K$_s$ data in W1 for sources with spectra close to that of an A0 star, and even deeper for 
moderately red sources like K stars or galaxies with old stellar populations.

The fundamental objective of WISE is to provide a sensitive all-sky 
survey in the mid-infrared. 
This survey will probably provide its biggest payoff in ways
that have not yet been imagined.   But several fields can be identified
where an all-sky survey will provide a guaranteed payoff.

Many classes of extragalactic objects will be quite red in the WISE colors,
as will brown dwarf stars which are among the Sun's closest neighbors.
Figure \ref{fig:bubble} shows the color-color diagram constructed from
the W1..W3 bands, with the regions occupied by various types of objects
illustrated.

\subsection{Brown Dwarf Stars}

\begin{figure}[t]
\plotone{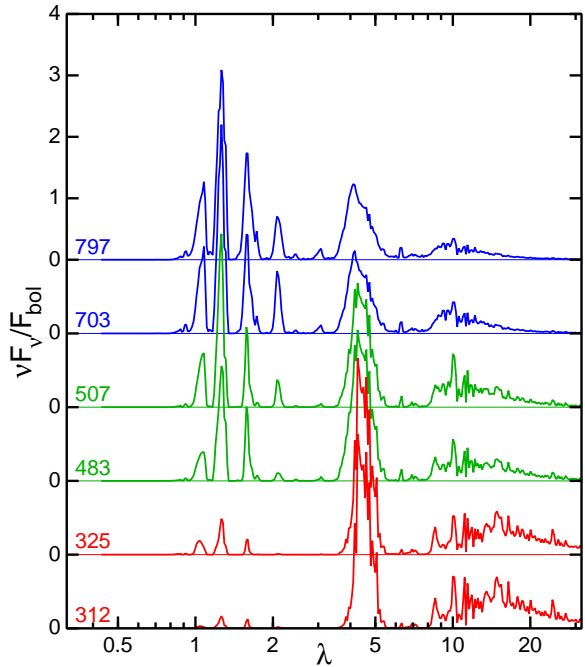}
\caption{Brown dwarf spectra selected from \citet{burrows/sudarsky/lunine:2003}
smoothed to 1\% resolution.  Curves are labeled with the effective temperature
from 797 K at the top to 312 K at the bottom.\label{fig:nuFnu_over_Fbol_vs_lambda}}
\end{figure}

\begin{figure}[tbp]
\plotone{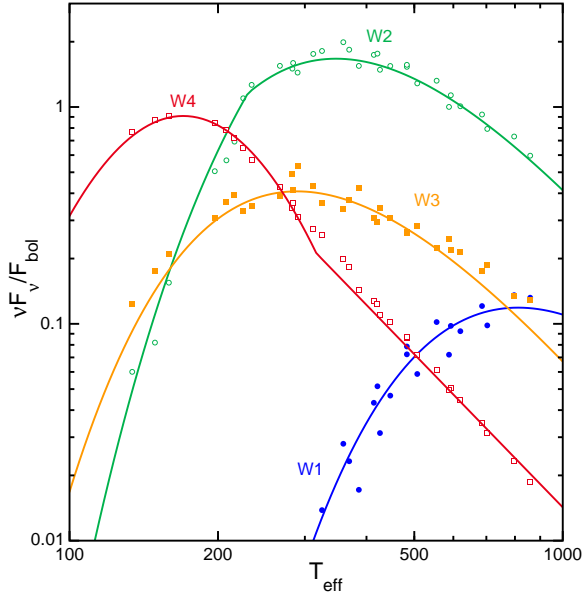}
\caption{Flux in the WISE bands compared to the bolometric flux
as a function of $T_{eff}$ for the models of \citet{burrows/sudarsky/lunine:2003}.
Simple functional fits are shown.
\label{fig:f-WISE-vs-Te}}
\end{figure}

Brown dwarf stars are very faint, since they are not massive enough
to fuse hydrogen into helium.  As a result, they gradually fade and cool,
and old brown dwarfs will be very cool and faint.  Both Jupiter with
$L = 10^{-9}\;L_\odot$ and Gliese 229B with $L = 10^{-5}\;L_\odot$
have very strong emission at 4.6 $\mu$m due to a lack of methane
absorption at this wavelength
\citep{kirkpatrick:2005}.  Thus the 4.6 $\mu$m band of WISE
is a powerful tool for finding cool brown dwarfs.  Most of the
brown dwarfs found prior to WISE have been discovered using the $z^\prime$
band of SDSS or the $J$ band of 2MASS or UKIDSS, but fairly high temperatures
are required before there is substantial emission in either of these
short wavelength bands.  As a result the currently known sample of brown dwarfs is biased
toward the hotter, and thus younger, objects.  WISE is able to
find 10 Gyr old brown dwarfs and thus should detect a high
density of stars in the solar neighborhood.   The expected number density
of brown dwarfs is 1-2 times the density of ordinary stars 
\citep{reid/etal:1999,chabrier:2002}.

The expected number of brown dwarfs that WISE will see can be
computed using models for the emitted spectra of brown dwarfs
and the luminosity and effective temperature
of brown dwarfs as a function of mass and age
\citep{burrows/sudarsky/lunine:2003}.  These model spectra
are plotted as $\nu F_\nu/F_{bol}$ in Figure 
\ref{fig:nuFnu_over_Fbol_vs_lambda}.
The spectra can also be used to compute the dimensionless
ratios of $\nu F_\nu/F_{bol}$ for each WISE band, which turn out to be
well described by simple functions of the effective
temperature $T_e$.
The functional forms plotted in Figure \ref{fig:f-WISE-vs-Te}
are given by:
\be
\frac{\nu F_\nu}{F_{bol}}  =  0.16 \frac{15}{\pi^4} \frac{x^4}{e^x-1}
\ee
with $x = h\nu/(1.34kT_e)$ for band 1 in blue,
\be
\frac{\nu F_\nu}{F_{bol}}  =  2.25 \; \mbox{min}(T_e/230,1)^2 \; \frac{15}{\pi^4} \frac{x^4}{e^x-1}
\ee
with $x = h\nu/(2.29kT_e)$ for band 2 in green,
\be
\frac{\nu F_\nu}{F_{bol}}  =  0.55 \frac{15}{\pi^4} \frac{x^4}{e^x-1}
\ee
with $x = h\nu/(1.05kT_e)$ for band 3 in orange, and
\be
\frac{\nu F_\nu}{F_{bol}}  =  0.9
\cases {\exp[-5.4\log(T_e/170)] & if $T_e > 317$ \cr
\exp[-20\log(T_e/170)^2] & otherwise}
\ee
for band 4 in red.
Thus WISE band 2 sees a flux for $T_{eff} > 230$ K that is similar to a greybody with a temperature
2.25 times higher than $T_{eff}$, and with an emissivity 2.25 times
higher than a blackbody.  The fluxes in WISE bands 1 and 3 also look like higher
temperature greybodies but with emissivities much smaller than a blackbody.
The flux in WISE band 4 does not behave like a blackbody.

\begin{deluxetable}{rrrrrr}
\tablecolumns{5}
\tablecaption{\label{tab:NBD}Number of brown dwarfs with fluxes greater than the WISE
4.6 $\mu$m required flux limit of 160 $\mu$Jy for either the \citet{chabrier:2003}
log normal IMF (lgnm) or \citet{reid/etal:1999} power laws with various indices $\alpha$
from 0.4 to 1.3, while suppressing a fraction $S$ of the 4.6 $\mu$m flux.}
\tablehead{
\colhead{IMF} & \colhead{$S$} & \colhead{$T_e < 300$} &
\colhead{$T_e < 500$} & \colhead{$T_e < 750$} & \colhead{$d < 1.3$~pc} 
}

\startdata
\hline
lgnm     & 0.000  &   7.1  & 228.5 & 1386.0 &  1.20 \\
\nodata & 0.489  &  1.5   &  55.0  & 400.0   &  1.19 \\
 & & & & & \\
 0.4       & 0.000  &   2.6  &  79.0  & 520.8   &  0.47 \\
\nodata & 0.125  &  1.9   &  58.6  & 400.0   & 0.47 \\
 & & & & & \\
 0.7       & 0.000  &  5.2   & 125.0 & 693.9   &  0.68 \\
\nodata & 0.251  &   2.7  &  67.6  & 400.0   &  0.67 \\
 & & & & & \\
 1.0       & 0.000  &  10.7 & 203.2 & 952.9   &  1.06 \\
\nodata & 0.375  &   3.8  &  76.9  & 400.0   &  1.03 \\
 & & & & & \\
 1.3       & 0.000  &  22.5 & 341.5 & 1356.1 &  1.82 \\
\nodata & 0.493  &  5.3   &  88.9  & 400.0   & 1.72 \\
\hline
\enddata
\end{deluxetable}

But when \citet{eisenhardt/etal:2010} surveyed 10 square degrees with Spitzer
to a flux limit at 4.5 $\mu$m 20 times deeper than the WISE all sky
sensitivity, they found fewer very red sources than predicted
by combining the \citet{burrows/sudarsky/lunine:2003} models
with either the \citet{chabrier:2003}
log normal IMF or the  \citet{reid/etal:1999}  power laws.  The models could
be brought into agreement with the observed counts by assuming that about
50\% of the IRAC 4.5 $\mu$m flux was suppressed by an absorber other than 
methane or water vapor.
The amount of absorption needed is highly correlated with the assumed mass
function, but the  \citet{eisenhardt/etal:2010} observation of  about 10 brown dwarfs
gives a model-independent prediction that WISE should find about
$10(41253/10)/20^{1.5} \approx 400$ late-T and Y class brown dwarfs.
Table \ref{tab:NBD} shows the number of brown dwarfs in various temperature
or distance bins that are detectable by WISE for different assumed mass functions.
The age distribution is assumed to be uniform from $10^8$ to $10^{10}$ years,
and the mass function is integrated from 1 to 80 $M_J$.
For each mass and age bin the luminosity and effective temperature are found by 
interpolation.   
A fraction $S$ of the 4.6 $\mu$m flux is then suppressed, 
following \citet{golimowski/etal:2004}, who found a suppression factor
$1/(1-S) = 1.5$ to 2.5, implying $S = 0.5 \pm 0.1$.
The volume in which the
object could be detected is computed, multiplied times the space density
for the mass and age bin, and summed over all masses and ages to give the
numbers in Table \ref{tab:NBD}.  There are two lines for each mass function:
the first does not include the W2 flux suppression, and the second has $S$ adjusted
to give the expected 400 detectable objects with $T_e < 750\;\mbox{K}$.  Matching the 
\citet{eisenhardt/etal:2010} source density with a flux suppression in the range recommended
by \citet{golimowski/etal:2004} requires a high number density of brown dwarfs, as
given by  \citet{chabrier:2003} or \citet{reid/etal:1999} with a fairly steep power law index.

The WISE sample of brown dwarfs will be 3 magnitudes brighter than the
Spitzer sample of  \citet{eisenhardt/etal:2010}, and thus much easier to follow up with
the HST or the Keck telescopes.  Astrometric followup will be needed to pin down
the properties of old, cold brown dwarfs.

\subsection{Ultra-Luminous Infrared Galaxies}

\begin{figure}[tbp]
\plotone{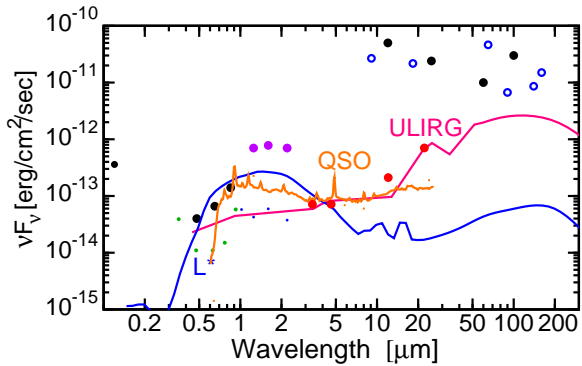}
\caption{Plot showing the spectrum of an $L_*$ galaxy at redshift $z = 0.33$,
a ULIRG (FSC15307 redshifted to $z=3$ and scaled by a factor of 3), and observations
of the $z = 6.42$ QSO J1148+5251 with a scaled composite quasar spectrum
in orange, compared to the WISE all-sky expected sensitivities in red.
Other surveys are indicated with dot areas corresponding to their
sky coverage: from the left,
GALEX AIS \citep{martin/etal:2003},
SDSS in green \citep{york/etal:2000},
DPOSS in black \citep{djorgovski/etal:1998},
UKIDDS LAS in blue \citep{lawrence/etal:2007},
2MASS in magenta \citep{skrutskie/etal:2006},
AKARI in blue open circles \citep{matsuhara/etal:2005},
and the IRAS FSC in black \citep{moshir/kopman/conrow:1992}.
\label{fig:Lstar-ULIRG-QSO}}
\end{figure}

\begin{figure}[tbp]
\plotone{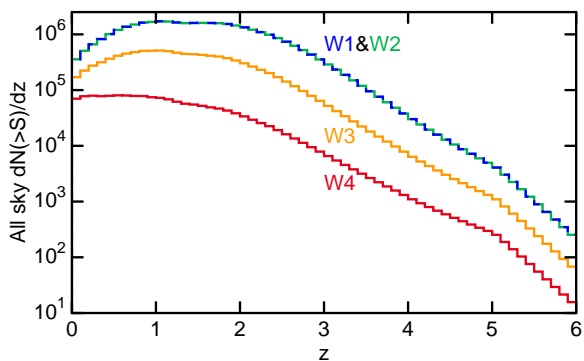}
\caption{Number of QSOs and AGN brighter than the
expected sensitivity in the 4 WISE bands over
the entire sky, based on the \citet{hopkins/richards/hernquist:2007}
luminosity function.
The W1 and W2 curves lie almost on top of each other,
with W3 and W4 giving lower counts.
\label{fig:dNgtSdz-W1234-sky}}
\end{figure}

WISE will detect the most luminous galaxies in the Universe.
The Ultra-Luminous Infrared Galaxies (ULIRGs) are due to mergers
that lead to dust-enshrouded star formation and also to AGN
activity as gas is disturbed out of stable circular orbits
and falls into the central super-massive black holes in the
merging galaxies.
Figure \ref{fig:Lstar-ULIRG-QSO} shows typical ULIRG and QSO
spectra compared to the WISE flux limits.

In the concordance $\Lambda$CDM cosmology structure formation
proceeds in a bottom-up manner, so large objects are formed by
the mergers of smaller objects.   The peak of large mergers occurs at redshifts 1 to 2 and
leads to the peak in the star formation history of the
Universe, and WISE has sufficient sensitivity at 22 $\mu$m
to see the top end of the luminosity function at redshift 3.
As a result, WISE will detect many galaxies with $L > 10^{13}\;L_\odot$.
Quasar activity is also associated with mergers, and the
evolving quasar luminosity function of \citet{hopkins/richards/hernquist:2007} 
predicts that WISE will see over
$1.5 \times 10^5$ quasars at 22 $\mu$m and over 3.6 million
quasars at 4.6 $\mu$m, with redshift distributions shown in
Figure \ref{fig:dNgtSdz-W1234-sky}.

The abundance of ULIRGs as a function of luminosity and redshift will
provide important information about the formation mechanisms that
have led to the galaxies and clusters we see at low redshift today.

\subsubsection{Red/Dusty AGN and QSOs}

WISE will characterize and probe fully the extent of the population of obscured 
active galactic nuclei and will provide a highly uniform set of mid-infrared 
photometry for virtually all known Active Galactic Nuclei (AGN) and 
Quasi-Stellar Objects (QSOs).
Most of what is known about the properties and evolution of AGN and QSOs 
has been deduced from studies of objects discovered in ultraviolet-excess and 
optical emission line surveys. However, infrared and radio surveys reveal 
that many AGN in the universe have remained hidden from short wavelength 
surveys because of reddening and obscuration by dust in and around their 
nuclei \citep{low/etal:1988,webster/etal:1995}. 
Estimates of the fraction of AGN missing from optical/UV selected samples vary considerably,
ranging from 15 to over 50\% 
\citep{richards/etal:2003,brown/etal:2006,glikman/etal:2004,glikman/etal:2007}.
Such an obscured population 
may account for at least part of the hard X-ray background \citep{comastri/etal:1995}, 
and it may contribute measurably to the far infrared background. 

The most extensive searches for red AGN have been based on the 2MASS PSC, using either
near-infrared color selection or combination with mid- and far-infrared and radio measurements
\citep{cutri/etal:2001,gregg/etal:2002,glikman/etal:2007}.  These studies have found large numbers
of predominantly low redshift AGN, adding considerably to the complete census of AGN
in the local universe.  However, 2MASS is a relatively shallow survey and is biased against the most heavily
obscured AGN at higher redshifts.  WISE will extend the 2MASS studies to higher redshifts, probing
the luminosity and number density evolution of red AGN over a significant fraction of the age of the
universe.

One of the highest redshift QSOs, SDSS 1148+5251, is shown in 
Figure \ref{fig:Lstar-ULIRG-QSO}, with data from \citet{jiang/etal:2006}
and a scaled composite spectrum constructed from 
\citet{vandenberk/etal:2001} and \citet{glikman/helfand/white:2006}.
A similar source is detectable by WISE at 3.4 \& 4.6 $\mu$m over the whole sky, and also at
12 $\mu$m with the better sensitivity achieved at the actual ecliptic latitude of this source
($\beta = 46^\circ$).

\subsection{Asteroids}

\begin{figure}[tbp]
\plotone{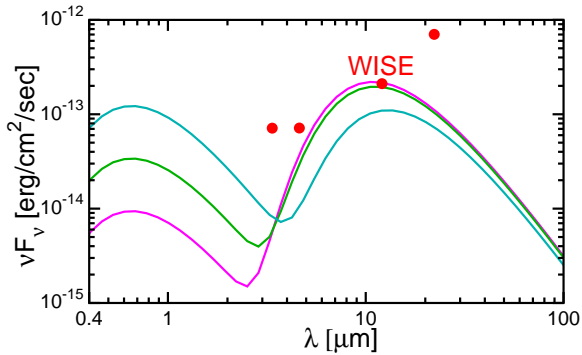}
\caption{Fluxes from an 0.13 km diameter near Earth asteroid
at a distance of 0.5 AU with elongation $90^\circ$ for three albedos:
0.04, 0.145 and 0.52 from bottom to top on the left side of the plot.  
These models assume a Lambertian
surface.  The phase angle is $63^\circ$.
The WISE  $5\sigma$ point source sensitivities for 8 stacked frames are
shown as solid circles.\label{fig:flux-vs-albedo-WISE}}
\end{figure}

Asteroids are typically fairly dark, but do have a large range
of albedos.  As a result, the visual magnitude is an imprecise
indicator of the size of an asteroid.   
\citet{stuart/binzel:2004} show a range in albedos from 0.023 to
0.63 for near Earth asteroids, leading to a 5:1 range in
estimated diameters and a 125:1 range in volumes.
But the light that is not
reflected is absorbed and reradiated in the thermal infrared.
Figure 
\ref{fig:flux-vs-albedo-WISE} shows fluxes from an asteroid with
low, median and high albedos.  
The ratio of optical to infrared fluxes gives the albedo,
and the infrared flux and color temperature give a good
indication of the diameter of an asteroid.

WISE is sensitive enough to see 1.2 km diameter
asteroids in the main belt at 2.7 AU from the Sun, and
0.13 km diameter near Earth objects that are 0.5 AU from
the Earth and 1.1 AU from the Sun.   These limits are
for objects with known orbits so that multiple frames
can be stacked, and are for 5$\sigma$.  This gives
10\% accuracy in diameters, which is comparable to
the errors caused by the unknown thermal inertia
and rotation poles of most asteroids \citep{wright:2007}.  
WISE will be able to provide
radiometric diameters for about $3 \times 10^5$
asteroids using image stacking analysis.
For an asteroid with a bolometric albedo
of $A = 5\%$, the WISE stacked frame sensitivity of 800 $\mu$Jy at 12 $\mu$m
corresponds to an optical flux of roughly
\begin{equation}
F_\nu(opt) \approx \frac{\nu_{IR}}{\nu_{opt}} \frac{A}{1-A} F_\nu(IR)
= 2\;\mu\mbox{Jy}
\end{equation}
which is an optical magnitude of 23.  If good optical data
are available, all of these asteroids will have well-determined 
albedos.

For new asteroids, a diameter of 2.4 km in the main
belt or 0.25 km for a Near Earth Object (NEO) at a distance of 0.5 AU
gives enough signal-to-noise
in a single frame to allow detections.   The NEOWISE program
\citep{mainzer/etal:2010}
is searching the WISE data for moving
objects and sending positional information to the IAU Minor Planet
Center.   WISE should detect
more than $10^5$ asteroids on single frames.  

WISE provides about 10 measurements of the flux spaced
over 30 hours for a typical asteroid
which can be used to derive rotation periods
for tens of thousands of asteroids.

The accurate diameters for previously known asteroids plus new asteroids 
derived by the WISE task will allow a significantly more precise knowledge 
of the size frequency distribution (SFD) of sub-km NEOs.  Since only a few 
dozen sub-km objects currently have diameters that have been accurately 
measured by either radar or thermal IR data, WISE data will significantly 
improve our knowledge of the SFD of these smaller objects.  Diameters derived 
from WISE measurements will reduce uncertainties in the SFD associated with 
the traditional reliance on the optical H magnitude.  Also, since infrared discoveries 
of new asteroids are relatively insensitive to albedo, the set of new discoveries made
by WISE are much closer to a diameter-limited survey than optical surveys, 
resulting in much better SFDs.
 
By observing asteroids at the time of their local afternoon and comparing this to their 
thermal fluxes in their local morning, WISE would enable measurement of the 
temperature differentials that give rise to the Yarkovsky effect, a phenomenon 
that is often the dominant uncertainty in the long term orbit propagation for small 
asteroids.
Measuring the Yarkovsky effect via the thermal infrared 
provides rather different information for a different population
than the radar method \citep{chesley/etal:2003}.  The radar 
technique uses precision observations of an asteroid's trajectory over a decade to look 
for small changes to the ephemeris caused by the Yarkovsky force.  In order to obtain 
sufficient astrometric precision, only NEOs which are very nearby can be observed.  
Radar data allow one to compute the tangential acceleration $a$ produced by the
Yarkovsky effect.
However, using infrared to look for temperature differentials between an asteroid's 
local morning and evening requires only two epochs of observation separated by 3-8 months.  
The infrared technique is not limited to NEOs and will also work for Main Belt asteroids such as the
Baptistina family. 
The infrared data allow one to compute the diameter which gives the mass $m$, and the
morning \vs\ evening difference gives the tangential force $F$.  
Thus combining IR and radar data allows measurements
of all the quantities in $F = ma$.
For asteroids in a family, the change in semi-major axis divided by the age of the family
gives an estimate for the acceleration $a$, so the IR data can be used to verify the time
since the collision that created the family.  The hypothesis \citep{bottke/vokrouhlicky/nesvorny:2007}
that the K/T impactor was part of the Baptistina family depends on the estimated
160 Myr age of this family, so measuring the Yarkovsky force and diameter for a sample
of Baptistina family members will provide a valuable test of this idea.
The infrared Yarkovsky detection technique is improved by knowledge of an objectÕs 
lightcurve, well measured by WISE.
\subsection{Other Science Goals}

An all-sky survey in the thermal infrared can be used to address
many scientific questions, not all of which require surveying the
whole sky.  But by surveying the whole sky, WISE will enable
these projects among many others:

\subsubsection{Comet Trails and Zodiacal Bands}

WISE will study the nearest planetary debris disk: our own zodiacal cloud 
and asteroid system.  The interplanetary dust cloud is a complex distribution 
of dust grains from numerous sources evolving under the influence of 
planetary perturbations, collisional evolution and solar radiation sources.  
The zodiacal dust cloud was long thought to come from comets 
\citep{whipple:1955,whipple:1967,whipple:1976} but current estimates of cometary dust 
production are an order of magnitude below that needed to maintain the 
cloud.  The IRAS dust bands show that asteroids are a substantial contributor 
to the interplanetary dust cloud.  There are at least seven dust band pairs 
girdling the ecliptic.  Comet debris trails were also discovered by IRAS 
\citep{sykes/etal:1986,sykes/lien/walker:1990} which provided more than 100 observations of the 
trails associated with 8 short period comets.  WISE scans through the 
ecliptic 30 times per day which will permit frequent observations of dust bands 
and comet trails with an angular resolution many times better than IRAS.  
This allows a detailed assessment of the population and evolution of these 
structures in the zodiacal cloud.

\subsubsection{Young Stars and Debris Disks}

\begin{figure}[tbp]
\plotone{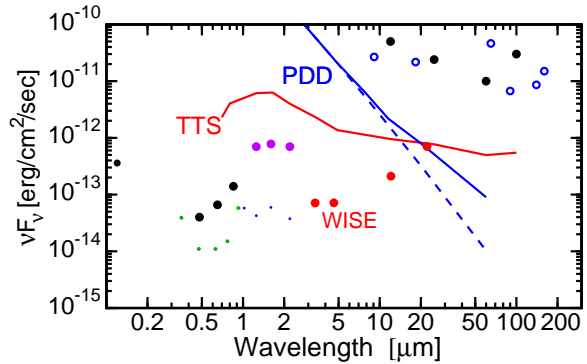}
\caption{Figure showing an M0 T Tauri star at a distance of 2.5 kpc (TTS)
and an A2 star with a planetary debris disk at 300 pc (PDD) compared to
the WISE expected sensitivities.  The unlabeled points for other surveys are
the same as in Figure \ref{fig:Lstar-ULIRG-QSO}.
\label{fig:TTS-PDD}}
\end{figure}

WISE will provide a robust statistical database for studying star formation, 
the evolution of circumstellar disks, and the dissipation of zodiacal clouds 
around stars of all masses, extending even into the brown dwarf mass regime. 
Detecting the small IR excess from the later stages of disk dissipation requires 
the ability to accurately predict the long wavelength photospheric emission and 
the ability to accurately measure the total long wavelength emission.  
The simultaneous 4 color photometry provided by WISE will enable the required 
predictions, and the sensitivity of WISE will enable detection of optically thick
disk emission for solar type stars out to 1 kpc distance and down to the hydrogen burning limit
in the Taurus and Ophiucus star forming regions. Because young stars are 
 variable, the simultaneous four-color coverage of WISE provides a precise 
 means of characterizing thermal excess from circumstellar disks or zodiacal 
 material. (A possible second coverage of the object by WISE 6 months after the 
 first will even provide clues as to the nature of the variability, e.g. extinction vs. 
 accretion.) 

Because WISE is an all-sky survey it will reveal isolated young stars well outside 
obvious star-forming regions. For example, two of the nearest clusters (of any age) 
were discovered only within the last decade though their YSO-like properties: 
the TW Hydrae Association (d = 55 pc, by infrared emission) and the Eta 
Chamaeleonis Cluster (d = 96 pc, by X-ray emission). Both have about a dozen 
known members and are younger than 10 Myr. Surprisingly neither is associated 
with a molecular cloud. TW Hya covers more than 700 sq. degrees. A complete 
survey would be an impossible task for missions like Spitzer but is the natural
 focus of WISE. Such intermediate-age very-nearby clusters provide an ideal 
 laboratory for studying the late stages of planetary accretion.

WISE will find young stars which are still accreting molecular cloud material over 
a substantial fraction of the Galaxy:
WISE will measure stars with optically-thick circumstellar disk emission down 
to the hydrogen-burning limit (0.08 M$_\odot$) out to 500 parsecs (e.g. the Orion 
molecular cloud). Solar-type ``Class I'' and T-Tauri stars with optically-thick disks 
will be found to substantial distances (Figure \ref{fig:TTS-PDD}
shows the M0 T Tauri star Sz 82 scaled to a distance of 2.5 kpc), thanks to their 
strong excesses at 12 and 22 $\mu$m. Expanding upon Spitzer's survey of 
selected regions of a few molecular clouds, WISE will provide a full inventory of 
low-luminosity young stellar objects in many dozens of star forming clouds 
extending over the entire sky. WISE will also discover hundreds of the very 
youngest stars: the elusive ``Class 0'' sources. These objects, which are in the 
early stages of gravitational collapse, are rare because they remain in this phase 
for less than 100,000 yr. They are deeply embedded in dust, so they are visible 
only at wavelengths longer than 20 $\mu$m, although they may be seen in scattered light
in the mid-IR.

WISE will survey planetary debris disks
around thousands of stars in the solar neighborhood.
During the early stages of planet formation micron-sized grains accrete 
into successively larger particles. The resulting reduction in grain surface 
area makes the circumstellar disk optically thin, and the infrared excess 
substantially decreases. The disk does not clear uniformly due to both the 
variation in accretion/dynamical timescale with distance from the star, and 
the shepherding effects of forming planets \citep{kuchner/holman:2003}. 
This non-uniformity is evident 
in the distribution of infrared excess as a function of wavelength through 
the mid-infrared. WISE will exploit these signatures to reveal the 
systematics of and refine the timescales for disk clearing and solar system 
formation.

Although the planetesimal/planet formation phase is short-lived, dust 
excess signatures can persist into the main-sequence phase as 
demonstrated in stars like Vega by IRAS \citep{aumann/etal:1984}. 
These excesses arise from small 
grains with Poynting-Robertson lifetimes of tens of thousands of years. Their 
existence argues that they must be actively replenished by the erosion of 
planetesimals and thus they are signatures of successful planetary formation. 
Indeed, high resolution 20 $\mu$m imaging of infrared-excess objects like HR4796 
reveal dust rings which may betray the existence of invisible planets.
Figure \ref{fig:TTS-PDD} also shows the known debris disk system $\zeta$ Lep
scaled to a distance of 300 pc.  With the short wavelength WISE bands pinning
down the photospheric flux, the excess at 22 $\mu$m is easily detectable.

The sensitivity to see the photospheric flux and enough wavelength coverage 
to predict the photometric flux are, once again, the key to revealing excesses 
from eroding planetesimals. In addition to studying the statistical distribution 
of such small IR excesses, of great interest to the TPF mission, WISE will 
provide JWST with hundreds of targets for high-resolution imaging akin to the 
ground-based and HST images of HR4796.

\subsubsection{Interstellar Dust}

WISE detects most components of interstellar dust and produces very good 
maps of interstellar dust in the Galaxy. The 3.4 and 12 $\mu$m filters include 
prominent PAH emission features, the 4.6 $\mu$m filter measures the continuum 
emission from very small grains, and the 22 $\mu$m filter sees both stochastic 
emission from small grains and the Wien tail of thermal emission from large grains. 
The WISE survey will map the high latitude dust to a sensitivity 60 times better 
than IRAS at 22 $\mu$m and 200 times better at 12 $\mu$m. For typical high latitude dust 
showing PAH emission the 3.4 $\mu$m map will provide a signal to noise ratio 
comparable in sensitivity to the 12 $\mu$m map. The WISE survey thus extends 
the mapping of high latitude dust to cover the entire sky, filling in the holes left by 
IRAS in the dimmest regions of sky, and extending the Spitzer legacy projects
GLIMPSE and MIPSGAL to the entire sky.  The 3.4 $\mu$m map will provide the first high 
sensitivity map of extended emission of the entire sky at this wavelength.

These extended emission maps will allow study of the composition and detailed 
structure of interstellar dust. Unique to WISE, a comparison of the 3.3 $\mu$m PAH 
feature with other PAH emission features in diffuse clouds can yield information 
on PAH composition. Results of the study of dust in the Galaxy can also be 
compared to dust emission detected by WISE in nearby galaxies.

\subsubsection{Galactic Structure}

Roughly half of the sky is dominated by the Milky Way's 400 billion stars, many 
of which are variable in the infrared.  With deep simultaneous near- and 
mid-infrared observations, unique in an all-sky survey, the 4-color WISE will 
address global issues of Galactic stellar structure and populations: the disk's 
warp and flare; the nature of the bar, Gould's Belt, ring, and bulge; a detailed 
census of evolved stars; and the pursuit of halo star-streams from globular 
clusters and unmerged Galactic dwarf satellites.  WISE complements Spitzer's 
Legacy Project, GLIMPSE, which surveyed $< 300$ square degrees of the 
inner plane mostly constrained to $|b|<1^\circ$.  WISE will cover all the Galactic plane, 
and secure critical off-plane starcounts to probe the vertical population structure 
in the arms and spurs, and scale heights of the central bar, molecular ring and 
bulge (disk contamination diminishes rapidly off the plane).  Although the stellar 
counts are confusion limited near the Galactic plane with the WISE beam size and 
high sensitivity, the color-magnitude diagrams are robust to confusion effects; 
and thus delineation of differing stellar populations is still viable in the Galactic 
plane.

Off plane, unconstrained by confusion, WISE offers unparalleled infrared survey 
depth, uniquely capable of all-sky detection of even normal halo stars.  The 
``Spaghetti Survey'' \citep{morrison/etal:2000} of the halo uses optical photometry 
and spectroscopy to identify distant G-/K-giants and kinematic substructure 
caused by the destruction of accreted satellites. WISE will aid the optical quest 
for extra-tidal K-giants, detecting such stars to 70 kpc at 3.4 $\mu$m and 45 kpc 
at 4.6 $\mu$m, far beyond the depth of 2MASS at 2.15 $\mu$m.  Comparing the 
optical spectra and WISE data on the quasi-continuum (3.4 $\mu$m) and CO 
fundamental (4.6 $\mu$m) in these cool stars to synthetic spectra will enable the 
recognition of those stars that belong to remnant stellar swarms from partly 
digested satellites. 

\subsubsection{Nearby Galaxies}

\begin{figure*}[tb]
\epsscale{2.1}
\plotone{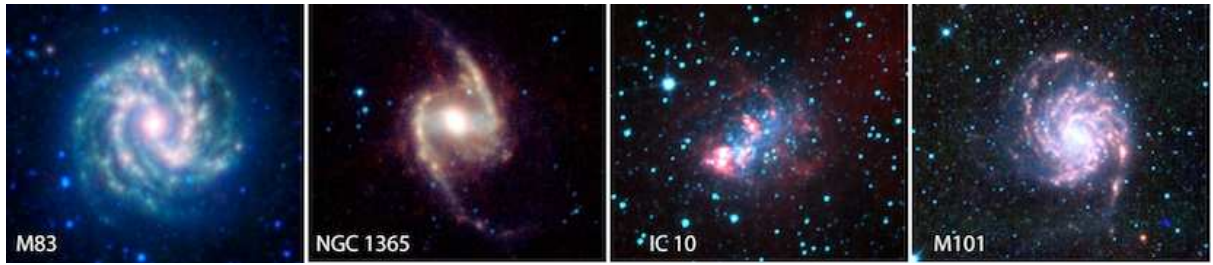}
\caption{Four nearby spiral and irregular galaxies that are well resolved by WISE.
\label{fig:nearby-galaxies}}
\end{figure*}

The angular resolution and sensitivity of WISE allow detailed study of the internal 
structure of galaxies in the local Universe, whose properties provide our basic 
understanding of high-redshift galaxies. For the $\sim 5000$ galaxies within 20 Mpc 
of the Sun, WISE spatial resolution is $< 500$ parsecs ($<1$~ kpc at 22 $\mu$m). 
Of this set, nearly 400 will be highly resolved, delineating globular clusters, giant 
molecular clouds and other discrete sites of star formation. 
Fig \ref{fig:nearby-galaxies} shows WISE images of 4 well resolved nearby galaxies.
The short mid-IR bands 
of WISE are an effective probe of the underlying stellar population, and are nearly 
immune to extinction.  The long bands are sensitive to hot dust associated with 
star formation, thus bridging the time span between evolving generations of stars. 
WISE data will complement the Spitzer Infrared Nearby Galaxies Survey Legacy project
\citep[SINGS,][]{kennicutt/etal:2004}, expanding the SINGS sample of 75 galaxies to many thousands
of galaxies, covering the widest range in morphological type.  It will
also supplement the
IRAC 3.6/4.5 $\mu$m Spitzer Survey of Stellar Structure in Galaxies \citep[S4G,][]{sheth/etal:2008}:
WISE will provide the crucial mid-IR tracers of star formation activity for
this complete sample of galaxies within 30 Mpc.

\subsubsection{Galaxy Clusters}

WISE will provide a census of star formation in very large samples of normal 
galaxies and clusters out to cosmological distances. 
WISE will detect over 100 million galaxies in at least one band, and 
many million galaxies in all four of its bands. 
The short wavelength W1-W2 color will be a useful redshift indicator for galaxies out to $z = 0.5$, 
and the flux measurements will 
provide an estimate for the total stellar mass of galaxies at $z < 0.5$ 
by measuring the rest frame 1.6-2.2 $\mu$m emission 
\citep{gavazzi/pierini/boselli:1996}.
WISE will thus provide a global history of star formation in massive field 
and cluster galaxies over the past 5 billion years. 

Cluster galaxies generally contain stellar populations that are old relative 
to those in the surrounding field galaxies, so the contrast of galaxy 
clusters relative to the field increases with redshift when viewed in the IR.  
By virtue of covering the entire sky, WISE will be able to detect 
all massive ($M > 10^{14} M_\odot$) galaxy clusters up to $z \sim 0.5$, 
enabling the most complete census of the stellar mass in dense environments 
when used in conjunction with large-area 
optical imaging surveys such as SDSS, PanStarrs, DES, and LSST. 
When used with the higher spatial resolution optical imaging from these surveys, 
which can help to overcome the confusion limit in the WISE images, 
WISE will also be able to identify even higher redshift clusters ($0.5 < z < 1$)  
over thousands of square degree areas at the poles where the larger than 
average number of WISE scans will lower the noise in the W1 data.  

\section{Data Products}

WISE will produce an image atlas in its 4 colors that
will be a stack of all the multiple frames covering each
part of the sky.  A catalog of sources extracted from
these images will also be produced.  These data products
are modeled on the 2MASS image atlas and the 2MASS
point source catalog.  WISE is not planning to produce a
separate extended source catalog.

In addition, the NEOWISE project is supporting the release 
of the individual WISE frames and a database of all the sources
extracted from the individual frames.  This dataset will be very
valuable for retrospective studies of asteroids discovered in
the future.

There are two data releases scheduled for WISE.  The first
release will be a preliminary catalog and image atlas based on early data
and covering more than 55\% of the sky, to be released
in April 2011.  The preliminary catalog will be
complete to a depth of 20$\sigma$.
A final data release covering the entire sky to a depth of
5$\sigma$ will be released 17 months after the last
data are taken.

\section{Prospects for Followup Observations}

The low background from space makes the WISE survey
very sensitive, and allows it to cover the largest possible solid
angle while still reaching fairly low flux levels.  The faintness
of the sources found by WISE makes followup with ground-based
telescopes quite difficult in the thermal infrared. 
However, the time required for
a detection scales like $B/D^4$ for background $B$ and telescope
diameter $D$, assuming diffraction-limited optics, so it will be possible
to confirm WISE sources with ground-based 10 meter class telescopes,
since the $(10/0.4)^4 = 390,625$ diameter factor almost compensates
for the factor of $2.5\times10^6$ in the backgrounds.  Since WISE
only integrates for 1.2 minutes on a typical source, 15 minutes on a
10 meter class telescope can surpass the WISE sensitivity on point
sources and give much better angular resolution.
Even though ground-based followup in the thermal infrared is possible, 
the most useful ground-based followups will probably be at
complementary wavelengths.  Examples include ground-based
$J$ band measurements of WISE brown dwarfs, ground-based optical
astrometry of WISE asteroids, ground-based optical spectroscopy
of WISE ULIRGs and AGN, and photometric redshifts of galaxies
based on optical and infrared colors.

The James Webb Space Telescope will also experience the low background intensity
seen by WISE, will cover the WISE passbands, and will be much larger than WISE.
Thus JWST will be the most capable platform for studying
WISE sources in the thermal infrared, since it will have both a large 
diameter and low background.   WISE data at 3.4 and 4.6 $\mu$m will be used
to improve the JWST guide star catalog by providing improved count rate
predictions and more recent astrometry.

\section{Conclusion}

WISE has exploited the low infrared background in space and the
power of large format infrared arrays  to map the entire sky with
unprecedented sensitivity and angular resolution between 3.4 and
22 $\mu$m.  The WISE catalog and image atlas will be extremely
valuable for a large number of astrophysical investigations ranging
from asteroid albedos in the solar system through the stellar
population of the solar neighborhood out to the most luminous star
forming galaxies in the visible Universe.

\acknowledgments

This publication makes use of data products from the Wide-field
Infrared Survey Explorer, which is a joint project of the University
of California, Los Angeles, and the Jet Propulsion Laboratory/California
Institute of Technology, funded by the National Aeronautics and
Space Administration.

\end{document}